\documentclass[twocolumn,twocolappendix]{aastex63}

\usepackage{xspace}
\usepackage[frozencache=true,cachedir=.]{minted}
\usepackage{graphicx}

\newcommand{\Ha}{H$\alpha$\xspace}
\newcommand{\Hb}{H$\beta$\xspace}
\newcommand{\NII}{[N{\sc ii}]\xspace}
\newcommand{\SII}{[S{\sc ii}]\xspace}
\newcommand{\OIII}{[O{\sc iii}]\xspace}
\newcommand{\OII}{[O{\sc ii}]\xspace}
\newcommand{\gleam}{\textsc{gleam}\xspace}
\newcommand{\gleamurl}{\url{https://github.com/multiwavelength/gleam}\xspace}

\received{December 5, 2020}
\revised{January 26, 2021}
\accepted{January 27, 2021}

\submitjournal{AJ}

\graphicspath{{./}{figures/}}

\usepackage{pgffor}
\usepackage[caption=false]{subfig}

\usepackage{amssymb}	
\usepackage{amstext}    
\usepackage{graphicx}

\begin{document}

\title{\gleam: Galaxy Line Emission \& Absorption Modeling}

\correspondingauthor{Andra Stroe}
\email{andra.stroe@cfa.harvard.edu}

\author[0000-0001-8322-4162]{Andra Stroe}
\altaffiliation{Clay Fellow}
\affiliation{Center for Astrophysics \text{\textbar} Harvard \& Smithsonian, 60 Garden St., Cambridge, MA 02138, USA}

\author[0000-0003-2606-4275]{Victor-Nicolae Savu}
\noaffiliation

\begin{abstract}
  We present \gleam (\textbf{G}alaxy \textbf{L}ine \textbf{E}mission \& \textbf{A}bsorption \textbf{M}odeling), a Python tool for fitting Gaussian models to emission and absorption lines in large samples of 1D extragalactic spectra. \gleam is tailored to work well in batch mode without much human interaction. With \gleam, users can uniformly process a variety of spectra, including galaxies and active galactic nuclei, in a wide range of instrument setups and signal-to-noise regimes. \gleam also takes advantage of multiprocessing capabilities to process spectra in parallel. With the goal of enabling reproducible workflows for its users, \gleam employs a small number of input files, including a central, user-friendly configuration in which fitting constraints can be defined for groups of spectra and overrides can be specified for edge cases. For each spectrum, \gleam produces a table containing measurements and error bars for the detected spectral lines and continuum, and upper limits for non-detections. For visual inspection and publishing, \gleam can also produce plots of the data with fitted lines overlaid. In the present paper, we describe \gleam's main features, the necessary inputs, expected outputs, and some example applications, including thorough tests on a large sample of optical/infra-red multi-object spectroscopic observations and integral field spectroscopic data. \gleam is developed as an open-source project hosted at \gleamurl and welcomes community contributions.
\end{abstract}


\section{Introduction}

One of the main goals of extragalactic astronomy is to understand the cosmic evolution of galaxies and black holes in the context of large scale structure. To obtain a comprehensive view of the physical processes driving their evolution and unveil their spatial distribution, spectroscopic observations of large samples of galaxies and active galactic nuclei (AGN) at increasingly high redshift are required. The most efficient way to obtain large samples ($>100$ to hundreds of thousands of sources) covering large volumes is through simultaneous observations of many objects. As a consequence, multi-object spectroscopy (MOS) and integral field unit (IFU) spectroscopy have experienced significant growth since the 1980s.

MOS plays an important role in repositioning mid-size telescopes, with instruments dedicated exclusively to completing large surveys of galaxies and quasars, such as LAMOST \citep{2012RAA....12.1197C}, WHT/WEAVE \citep{2012SPIE.8446E..0PD}, SDSS-IV \citep{2017AJ....154...28B} and DESI \citep{2016arXiv161100036D}. The instrument suite of $6-10$-m class optical/infrared telescopes usually contains MOS and IFU capabilities, e.g.\ Keck/DEIMOS \citep{2003SPIE.4841.1657F}, Keck/MOSFIRE \citep{2012SPIE.8446E..0JM}, VLT/VIMOS \citep{2003SPIE.4841.1670L}, VLT/KMOS \citep{Sharples2006}, Gemini/GMOS \citep{2004PASP..116..425H}, Subaru/FMOS \citep{2010PASJ...62.1135K}, MMT/Hectospec \citep{2005PASP..117.1411F}, Magellan/IMACS \citep{2011PASP..123..288D}. In the near future, new wide-field ($>1^\circ$), high-multiplex ($>1000$ targets) MOS instruments will be mounted, such as VLT/MOONS \citep{2012SPIE.8446E..0SC} and VISTA/4MOST \citep{2012SPIE.8446E..0TD}. All new-generation ground-based optical/infrared telescopes have MOS and IFU instruments planned, e.g.\ ELT/MOSAIC \citep{10.1117/12.2314135}, TMT/WFOS \citep{10.1117/12.672712} and GMT/GMACS \citep{2020SPIE11203E..08P}. Instruments on the flagship James Webb Space Telescope, including NIRSpec and MIRI, will also have MOS/IFU capabilities. MOS/IFU techniques have been also routinely used in the radio and sub-mm regime \citep[e.g.\ VLA, ALMA,][]{1980ApJS...44..151T, 2009IEEEP..97.1463W}.

As a result of transformational large-scale public surveys and concerted guaranteed time efforts completed over the past 3 decades, a growing body of spectroscopic observations have been made available to the community. Complementing guaranteed time observations and large scale public surveys, individual investigators have added MOS and IFU observations tailored to specific extragalactic science goals. Obtaining statistically robust samples and particular science cases that involve targets distributed across the sky (e.g.\ galaxy population studies in galaxy clusters, quasar surveys, high redshift galaxy surveys, or intra/circum galactic medium absorption-line surveys) require the combination of data coming from different telescopes and instruments. Further, the advent of online databases has made access to fully or partially reduced observations easier \citep{2019AJ....157...98G}, enabling individual authors to make use of existing spectroscopic observations for new science goals, possibly combining data from different telescopes. The sheer volume of data warrants automated analysis pipelines with minimal human interaction.

Striving for reproducible results, many authors in the field provide machine-readable data plus scripts used to obtain the results with the publication e.g.\ the Jupyter \citep{jupyter} notebook used for creating figures or the CASA script used to reduce and make images from ALMA data. However, publications mainly focus on the originally intended science case, leaving byproducts and intermediate results largely unreported (e.g.\ unreported line fluxes when the goal was measuring redshifts). In order to incorporate archival spectroscopic observations in new projects, researchers need to partially reproduce and build upon the efforts of the original authors.

The first fundamental property encoded by a spectrum is the source's redshift. A number of powerful, modern tools assist astronomers in obtaining accurate redshifts for large samples in an automatic and unsupervised way while also ensuring the reliability of the results \citep[e.g. EZ,][]{Garilli2010}. Apart from providing redshifts, the scientific potential of MOS and IFU observations is realized in extracting the (resolved) physics and chemistry of extragalactic objects from emission and absorption lines. With a growing body of literature with tailored science goals, each publication uses heterogeneous data and methods to measure emission and absorption lines. As the astronomy community further adopts the Python programming language \citep[e.g.\ Astropy,][]{2013A&A...558A..33A}, various interfaces for fitting functions exist. However, the low-level function fitting packages require individual authors to write their own bindings to interface between the reduced astronomical data and the fitting software. With a high level of duplicated effort in the community to write tailored code to fit spectral lines and with the high costs associated with sharing and maintaining it, access to data analysis software entails a great deal of overhead and represents a barrier to entry for the field of spectroscopy.

\gleam\footnote{\gleamurl} \citep[\textbf{G}alaxy \textbf{L}ine \textbf{E}mission \& \textbf{A}bsorption \textbf{M}odeling,][]{stroe_gleam} is a software tool for fitting Gaussian models to emission and absorption lines in large samples of galaxy and AGN spectra. \gleam has versatile science applications involving large samples of 1D spectra or IFU observations. For example, \gleam is ideally suited for unveiling the detailed physics and chemistry of galaxies, as derived from interstellar medium line ratios and stellar absorption lines, for a variety of samples spanning both cosmic time and environment. \gleam can also aid in the exploration of IFU cubes through spatially resolved physics, kinematics, and chemistry. Requiring only the source redshifts and with little to no interaction, the user can analyze large numbers of spectra in a uniform manner, even with data taken in different conditions, with different instrument setups, on different telescopes, at a range of signal-to-noise (S/N) regimes, and for a wide variety of sources. We tested \gleam mainly on optical and infrared spectra; however, we expect it to also work well on radio and sub-mm spectra.

In this paper, we provide an introduction to the \gleam software, focusing on features contained in the v1.0 release. Section~\ref{sec:gleam} describes the basic functionality, while Section~\ref{sec:software} discusses the necessary input files and expected outputs. Section~\ref{sec:examples} covers some example applications and uses. In Section~\ref{sec:development}, we present the open-source development model adopted for \gleam, while in Section~\ref{sec:future} we discuss possible extensions to the code in the near future.

\section{\gleam: Galaxy Line Emission and Absorption Modeling}\label{sec:gleam}

With \gleam, the user can process large numbers of sources in batch mode, taking advantage of the multiprocessing capabilities of modern CPUs. Optionally, \gleam also provides an interactive interface to inspect individual line fits on a spectral line and source-by-source basis. \gleam fits emission and absorption lines in fully-reduced 1D spectra using per-source spectroscopic redshift information and fitting constraints from a central configuration. The central configuration encourages users to define common fitting constraints for broad groups of spectra and be deliberate when defining overrides, which helps prevent user errors and facilitates an easier review of the methods and results by collaborators, referees, and readers. \gleam fits all lines listed in a central line list and can jointly fit lines located close together. \gleam also reports upper limits and identifies lines without spectral coverage. If so required, the user can provide a file containing sky bands, sky lines and/or OH lines to be masked and disregarded during line fitting. At its core, \gleam uses the popular LMFIT Python package\footnote{\url{https://lmfit.github.io/lmfit-py/}}\citep{lmfit} to perform the line fitting and to calculate and report errors on fit parameters. \gleam is also well integrated with Astropy\footnote{\url{https://www.astropy.org/}}\citep{2013A&A...558A..33A}, which enables the use of units and FITS tables.

As output, \gleam creates a FITS table with Gaussian line measurements and upper limits (as the case may be), including central wavelength, width, height, and amplitude, as well as estimates for the continuum under the line, the line flux, luminosity, equivalent width, and velocity width. \gleam can also make plots of the entire spectrum with fitted lines overlaid, as well as plots for each individual line fitted, using Matplotlib \citep{matplotlib}.

\gleam follows open-source practices, with planned features to be added to the living codebase published online on Github at \gleamurl. The latest release of \gleam can be installed easily with Python pip.

\section{The Software}\label{sec:software}

Below, we introduce \gleam's main functionality, features, required inputs, and outputs. For a full documentation of the code, we encourage the reader to consult \gleam's Github page at \gleamurl.

\subsection{Model fitting}\label{subsec:modelfitting}

In fitting the spectrum, \gleam groups neighboring spectral lines. For each spectral line group, a user-defined window of the spectrum around the group is considered for fitting. \gleam models each group as the sum of a constant for the continuum and one Gaussian for each spectral line. The assumption that the continuum is locally constant might fail if the window is too wide, while a too narrow window will not have enough line-free spectrum to properly constrain the continuum. Sections of the spectrum suffering from contamination, such as areas with sky lines, can also be masked.

The centers of all Gaussian components can be fixed, constrained to user-defined intervals, or left as free parameters. An initial guess for the central wavelength of each Gaussian is used to identify the component. This initial guess is calculated from the user-provided redshift for each spectrum and the global list of lines at rest-frame wavelengths. To offer the flexibility to fit emission and absorption lines in a range of galaxy and AGN spectra, \gleam relies on a single prior, the source redshift, for initializing the line fitting solution. In all but the brightest sources with the highest S/N spectral lines, a spectroscopic-quality redshift is required.

A fit is accepted when every Gaussian component passes the user-specified S/N ratio. When, due to noise and the overlap with sky lines, there is insufficient information in the data to fit the entire model, \gleam iteratively removes Gaussian components in search of an acceptable fit. Any removed Gaussian components are treated as non-detections and upper limits are computed for them.

\gleam employs LMFIT to perform the fitting \citep{lmfit}. Through a non-linear least-squares minimization using the Levenberg-Marquardt method, LMFIT enables the robust estimation of both model parameters and their errors.

\subsection{Naming convention}

When handling large numbers of spectra coming from different observations, sometimes from different telescopes, it is important to adopt a consistent naming convention. \gleam helps with this by prescribing a four-part hierarchical naming convention that allows for easy identification and grouping of spectra.

Each measured spectrum is uniquely identified by the combination of the following four properties:
\begin{description}
  \item[Sample] A label for the parent sample for the source, e.g.\ name of the parent galaxy cluster or famous field,
  \item[Setup] A label for the telescope, instrument, or mode used for the observation,
  \item[Pointing] An identifier for the individual pointing, fiber configurations, or slit configuration/mask the observation is part of,
  \item[SourceNumber] A source number to distinguish a target within a sample, setup, and pointing combination.
\end{description}

\subsection{Inputs}

\gleam uses five kinds of input files to gather information about spectra in order to compute the properties of its spectral lines:

\begin{itemize}
  \item A set of 1D spectra,
  \item Metafiles that provide a reference redshift for each spectrum,
  \item Configuration file, which specifies choices of spectral lines, fitting parameters, cosmological parameters and sky masking,
  \item Line table with the rest-frame wavelengths of the spectral lines of interest,
  \item (Optional) Sky band catalog, with details of any wavelengths contaminated by sky absorption/emission.
\end{itemize}

In a single run, \gleam can process spectra that originate from different data sets and which might have different units for wavelength or flux. It is therefore highly recommended to include units in the headers of all spectra files. \gleam propagates the units to the results. Line files and sky band files should also specify units in the file headers to ensure alignment with the spectra.

The metadata file contains information about individual spectra in the project, such as the setup and pointing they were observed with, a numeric identifier, and their redshift. The user may add custom columns to the metadata file to store other information about the spectra, such as the sky coordinates, quality flags, source types, etc. The project can have a single metadata file or multiple ones, as long as spectra are uniquely labeled. Because \gleam does not process the sky coordinates for sources, it cannot detect when two spectra pertain to the same source and, therefore, will produce separate independent fits for each input spectrum. It is incumbent upon the user to reason about which spectrum best fits their science requirements. When it is appropriate, another approach would be to combine/stack the relevant observations into a single spectrum before running \gleam.

\gleam can uniformly process large numbers of spectra, even with data taken in different conditions, with different instruments on different telescopes, and for a wide variety of sources. The configuration file is used to concisely describe how the different spectra should be processed, so they can be analyzed together. For easy editing and review, the configuration file for \gleam uses the YAML\footnote{\url{https://yaml.org/}} format. Taking advantage of the naming convention and the many reasonable defaults, the user can tailor the analysis at 3 levels. The global level parameters override the default configuration for all the spectra. The setup level offers a way to apply configuration overrides to groups of spectra (named setups). This level can be used to capture differences between telescopes or instruments, such as the spectral resolution. At the most granular level, the user can customize parameters for individual sources. While per-source overrides can help account for some particular cases (e.g.\ a small percentage of sources with both narrow and broad emission lines), they should be used sporadically due to the associated typing burden, and in the spirit of keeping the results comparable. The model parameters for each spectrum are computed by stacking the applicable overrides on top of the default in order: first, the global overrides, then any applicable per-setup overrides, and, finally, any applicable per-source overrides.

\gleam cannot specify any default for the line table and the instrumental resolution, so this information needs to appear at some level in the configuration file. With these two fields, we present a minimal working \gleam configuration example:

\begin{minted}{yaml}
globals:
  line_table: line_lists/Main_optical_lines.fits
  resolution: 4.4 Angstrom
\end{minted}

\gleam fits all lines listed in the line table and iteratively eliminates model components when the data does not yield satisfactory fits for all the lines. A S/N parameter, which defines the minimum accepted ratio between the estimated amplitude of a component and its error, separates detections from upper limits for each spectral line. In some setups, the user may select a starting subset of the lines and avoid unnecessary trials (e.g.\ excluding faint lines the user does not expect to be detected).

\begin{figure*}[ht!]
  \subfloat[A spectrum of a source at $z\sim0.1$, with a spectrum dominated by AGN features, including broad emission lines. The spectrum was taken with MMT/Hectospec, a multi-fiber instrument.]{\includegraphics[height=0.4\textwidth]{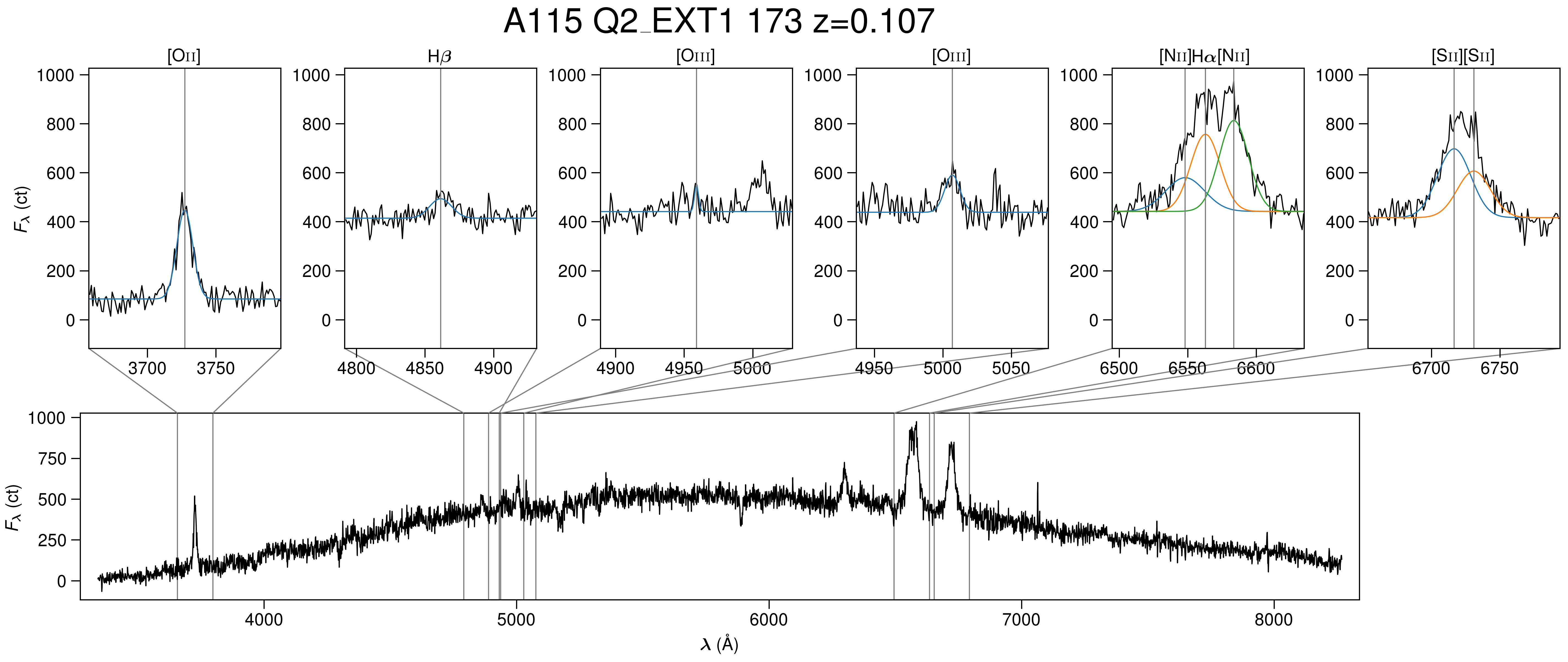}}\\
  \subfloat[A high S/N spectrum at $z\sim0.6$ dominated by star formation, taken in $>1''$ seeing conditions with the fiber-bed WHT/AF2 instrument.]{\includegraphics[height=0.4\textwidth]{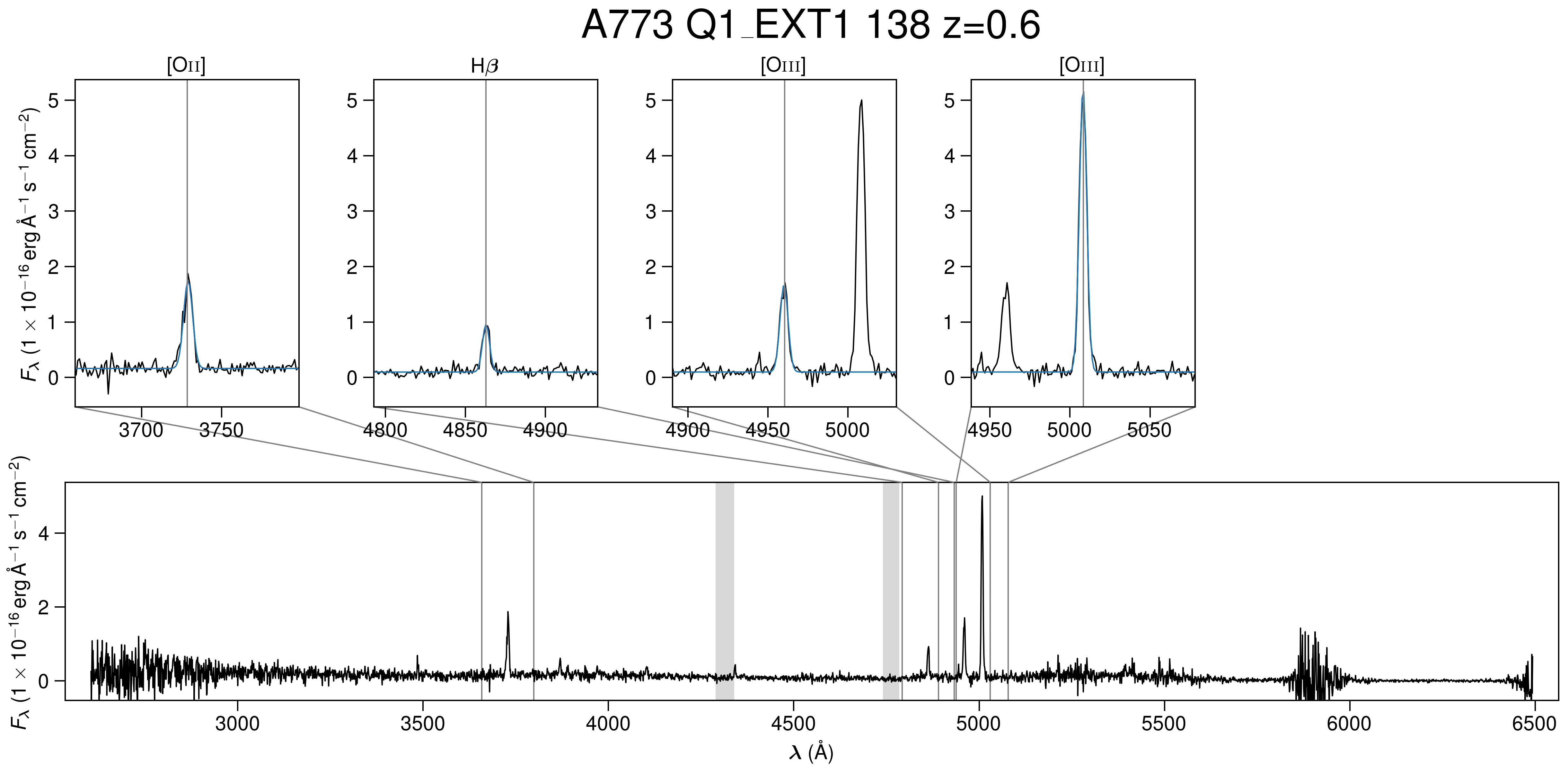}}
  \caption{\gleam emission and absorption line fits, highlighting different source types and origin telescopes.}
  \label{fig:spectra}
\end{figure*}

\begin{figure*}[ht!]
  \ContinuedFloat
  \subfloat[A star-formation dominated galaxy at $z\sim0.17$, with emission lines slightly overlapping with a sky-contaminated wavelength range. Data was taken in $\sim1''$ seeing conditions with thin clouds, with a multi-slit spectrograph (VLT/VIMOS).]{\includegraphics[height=0.4\textwidth]{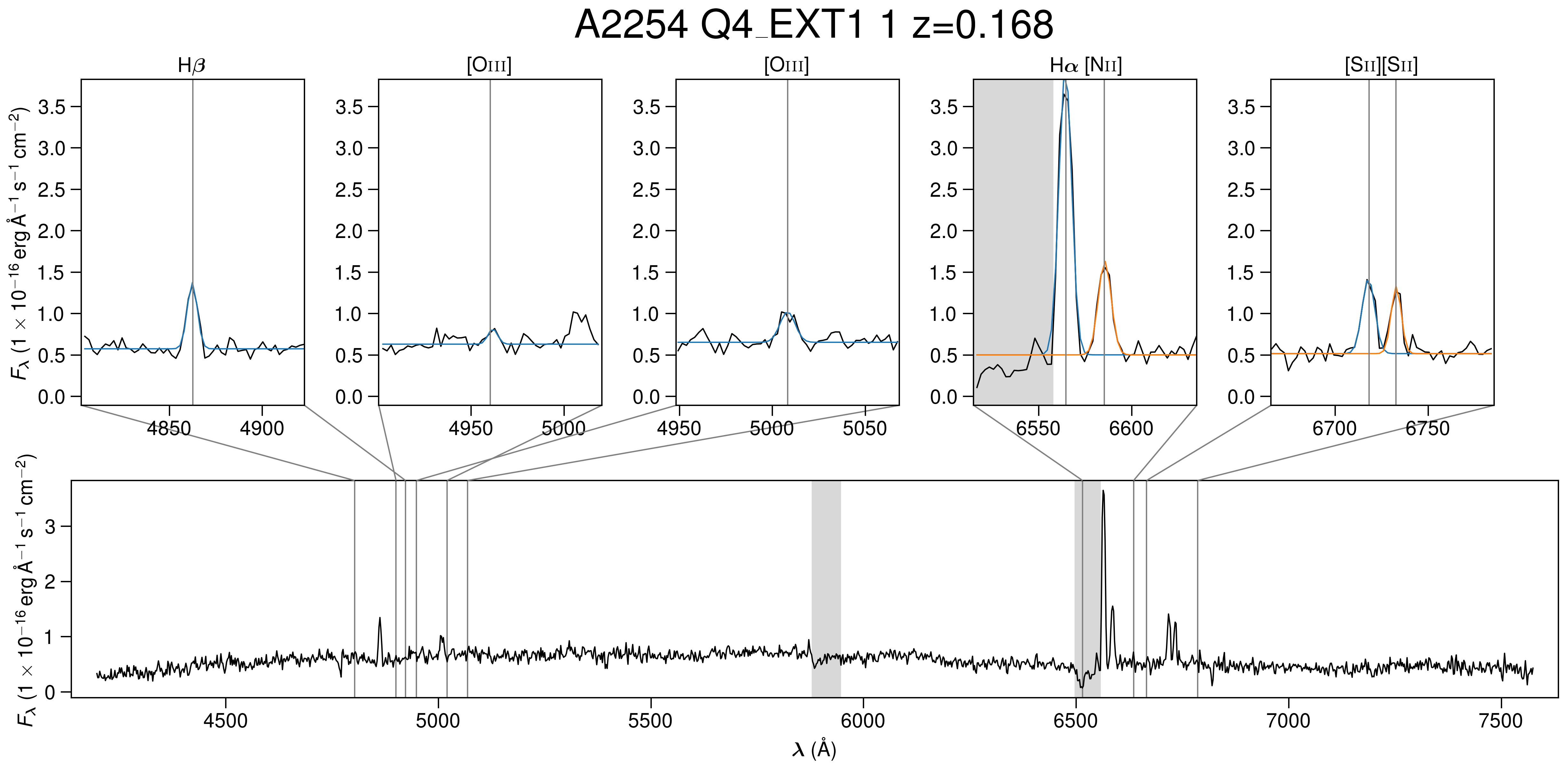}}\\
  \subfloat[A passive galaxy spectrum with absorption and emission features at $z\sim0.26$. Data was taken in $\sim1''$ seeing with MMT/Hectorspec instrument.]{\includegraphics[height=0.4\textwidth]{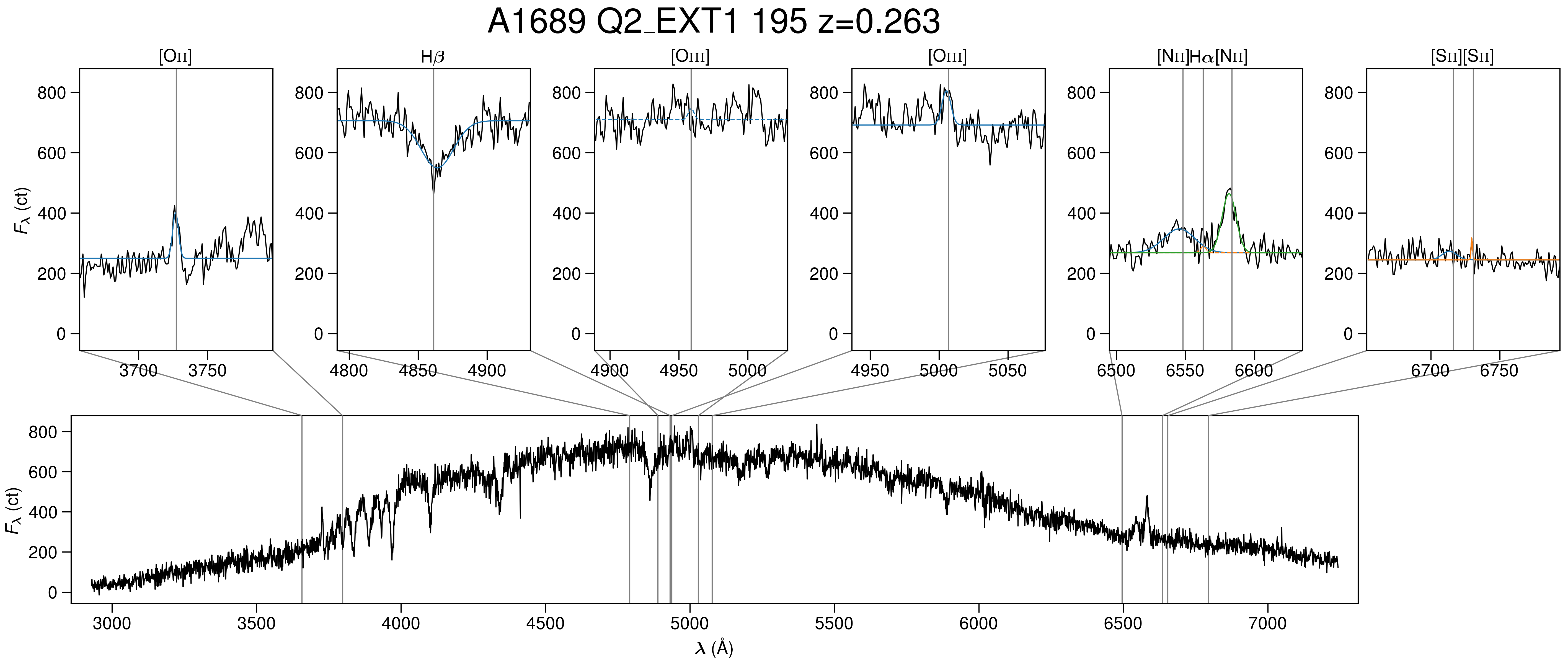}}
  \caption{Continued.}
\end{figure*}

By default, there is no sky masking and the entire spectrum is used. However, the user may control whether the model should ignore portions of the spectrum where sky bands may not have been reliably subtracted. These bands are masked and disregarded for fitting and treated as if no spectral coverage is available. For example, for a data set where sky subtraction only failed for a few of the spectra, the configuration may specify the sky band catalog at the global level but only turn masking on for individual sources (or setups) for which the sky subtraction is inadequate.

\gleam offers users a lot of flexibility in choosing the way line models are fit to the data. Neighboring Gaussian components can be fit jointly to account for nearby or blended lines. The user can also define the amount of continuum to be fitted on either side of a group, which should be large enough to encompass enough line-free continuum. Over the range specified, the continuum should be well approximated by a constant. Any unrelated lines that fall within the selected continuum are automatically masked. The center of each Gaussian component is first estimated based on the rest-frame wavelength in the line catalog and on the redshift estimate of the spectrum (listed in the metadata file). The Gaussian center can be fixed to the initial guess, allowed to vary within a small range around it, or can vary freely within the spectral range of data considered when fitting. This final option can lead to lines being mislabeled or cross-labeled, so it should only be used when the redshift estimate for the spectrum is so poor that neither of the two other options is feasible.

To report luminosities based on the fitted models, \gleam uses a set of cosmological parameters: Hubble constant ($H_0$), $\Omega_0$ and the CMB temperature. While the default values for these parameters are reasonably accurate and up to date, some projects may require slightly different values. The cosmology section overrides one or more of these parameters, and the resulting cosmology is then used consistently across all the spectra within a project. This is the only set of overrides that cannot be made on a per-setup or per-source basis, since doing so could produce results that are not comparable between sources.

\subsection{Outputs}

For each of the sources in the sample, \gleam produces a FITS table with all of the line fits and upper limits (with units derived from the input data, if available). Each line fitted is represented in a separate row, with all the corresponding line fit details contained in different columns. The output table contains fit parameters and their associated errors (such as the continuum estimation, central line wavelength, Gaussian height, standard deviation, and amplitude), line fluxes, luminosities, and equivalent widths. Each row also contains a flag for spectral coverage (i.e. whether the line is covered by the input spectrum) and another to indicate detection (whether the line is detected above the required S/N). Fit values and errors are omitted when the spectrum does not cover the spectral line. If a line is not detected, \gleam only reports an upper limit in the amplitude column and omits all other Gaussian fit parameters. The deconvolved full-width-at-half-maximum (FWHM) and the velocity FWHM are only reported if the line is spectrally resolved.

If plotting is enabled, \gleam produces two types of figures. The first type of figure shows the entire spectrum with zoom-ins on the emission and absorption line fits (see Figure~\ref{fig:spectra}). The second type of plot is focused on each line fit. Masked sky areas are shaded gray for clarity.

\section{Example applications}\label{sec:examples}

We demonstrate \gleam's capabilities by showcasing two natural applications, which also served as testbeds during the code development.

\subsection{A Large, Heterogeneous Sample of Extragalactic Spectra}

\gleam is well-suited for measuring emission and absorption lines in large, heterogeneous samples of extragalactic 1D spectra. In \citet{Stroe2021}, we thoroughly tested \gleam on all the spectroscopy available to us, which included about 4200 passive galaxies, star-forming galaxies, AGN, and quasars at redshifts from 0 to $\sim1$. The data were taken with four different instruments (VLT/VIMOS, WHT/AF2, Keck/DEIMOS, and MMT/Hectospec) that employ different techniques to achieve MOS capabilities (slits, fibers), under a range of sky, weather, and seeing conditions. In this project, the focus was on measuring optical emission lines, such as \OII ($\lambda\,3728$\,\AA), \Hb, \OIII ($\lambda\lambda\,4960,\, 5007$\,\AA), \Ha, \NII ($\lambda\lambda\,6550,\,6585$\,\AA), \Hb, and \SII ($\lambda\lambda\,6718,\,6733$\,\AA), with the spectral resolution for all instruments being sufficient for separating nearby narrow emission lines.

For the entire sample of 4200 sources, a simple, short configuration file was sufficient, as illustrated below. The fitting constraints, the set of spectral lines to be fit, and the sky lines to be masked were the same for the bulk of the sources. With \gleam, it was easy to set a different resolution for each instrumental setup and, when necessary, add, for example, a different continuum width (which resulted in more stable fits), turn off sky masking (when data reduction adequately corrected for the sky absorption), or use a different line table (when an air versus a vacuum wavelength calibration was applied to the data). We also set overrides for several individual sources. For example, we fit the full line list when \NII was also present in the data or when lines were blended. The YAML configuration file for this example application can be found below and demonstrates all the customization options that \gleam provides. Examples of line fits can be found in Figure~\ref{fig:spectra}.

\begin{minted}{yaml}
globals:
  sky: line_lists/Sky_bands.fits
  mask_sky: True
  line_table: line_lists/Main_optical_lines.fits
  lines:
    - OII
    - Hb 
    - OIII4
    - OIII5  
    - Ha
    - NII1
    - SII1
    - SII2
  fitting:
    SN_limit: 2 
    tolerance: 26.0 Angstrom 
    w: 3.0 Angstrom 
    mask_width: 20.0 Angstrom
    cont_width: 70.0 Angstrom
    center: constrained

setups:
  VIMOS:
    resolution: 12.5 Angstrom
    fitting:
      cont_width: 60 Angstrom
  MMT:
    resolution: 6 Angstrom
    line_table: line_lists/rsvao.fits
    mask_sky: False
  Keck:
    resolution: 1 Angstrom
    fitting:
      cont_width: 40 Angstrom
  WHT_R316R:
    resolution: 8.1 Angstrom
  WHT_R600R:
    resolution: 4.4 Angstrom

sources:
    ...
  A115.MMT.Q1_EXT1.194:
    lines: all
    ...
\end{minted}

Without plotting, the fitting for 10 spectral lines (with \Ha+\NII and the \SII doublet fit jointly) for the sample of $>4000$ sources could be completed in less 20\,min, using 6 threads on a modern laptop with a 2.9 GHz 6-Core Intel Core i9 processor. With plotting, the process can take up to 3\,h.

\subsection{Integral Field Unit Observations}

Very powerful Python wrappers tailored for the analysis of IFU data exist in the literature \citep[e.g.\ GIST,][]{2019A&A...628A.117B}. Their design is tailored to accomplish complex applications, such as the detailed modeling of absorption lines in passive galaxies, which require input spectra with good continuum detections. \gleam does not make assumptions on the underlying physics of the sources, making it a complementary tool for emission-line dominated sources that do not benefit from high S/N continuum detections.

In \citet{Stroe2020}, \gleam was used to measure spectral lines in Gemini/GMOS IFU observations of 5 emission-line dominated cluster galaxies at $z\sim0.2$. For analysis with \gleam, the IFU cube for each of the sources was split into individual spaxels. The nature of the project required slight reinterpretations of the naming convention. The \texttt{Sample} components was set to the name of the parent galaxy cluster, \texttt{Pointing} was used to specify the galaxy, while \texttt{SourceNumber} was used to label each spaxel in the IFU. Coordinates for each spaxel were added to the metadata file to track the connection between the spaxels and their sky positions with respect to each galaxy. This is a good example of using the metadata file for storing more than just the redshift information. With this interpretation of the naming convention, the YAML configuration for the project could be specified in just a few lines:

\begin{minted}{yaml}
globals:
  mask_sky: False 
  line_table: line_lists/rsvao.fits
  lines:
    - Ha
    - NII1
    - SII1
    - SII2
  fitting:
    SN_limit: 3 
    tolerance: 26.0 Angstrom
    w: 9.0 Angstrom
    mask_width: 20.0 Angstrom
    cont_width: 70.0 Angstrom
    center: constrained

setups:
  GMOS:
    resolution: 11.4 Angstrom
\end{minted}

\section{Development model}\label{sec:development}

\gleam is developed as an open-source project hosted at \gleamurl and published under the permissive BSD-3-Clause License. The authors welcome community contributions in the form of bug reports and feature suggestions, as well as code contributions under the same license via the GitHub pull-request system. \gleam is written in the Python programming language, which is a popular choice both within and outside the astronomical community, with the hope that interested contributors would find it easy to get started. The project aims to offer an inclusive and welcoming place for collaboration and has adopted the Astropy Community Code of Conduct\footnote{\url{https://www.astropy.org/code_of_conduct.html}}.

\section{Future developments}\label{sec:future}

In the near future, we will explore a number of natural extensions to \gleam.

In its first iteration, \gleam was envisioned to work out-of-the-box for most extragalactic science cases, hence the choice of Gaussian models for the fitting. In the future, we plan to expand the choices for component types with other models, such as Voigt (e.g.\ absorption lines towards quasars) or asymmetric (e.g.\ Ly$\alpha$ emission) profiles and more complex continuum models.

We aim to also provide better support for multiple component fits, such as when both broad and narrow lines are present in an AGN spectrum. In its iterative refinement of the model, \gleam removes Gaussian components in a sequential fashion. As such, the most complex model that converges is passed through the S/N criterion to identify detected spectral lines. As evidenced in Section~\ref{sec:examples}, \gleam robustly fits well-separated spectral lines at non-redundant wavelength separations. For scenarios in which many ($>3$) blended or nearby lines at lower S/N are jointly fit, sometimes lines are cross-identified. Incorrect matching/labeling of spectral lines can be avoided by making use of constraints on the center of each Gaussian component. In the future, a number of new additions to \gleam will ensure successful and correct fits to a wider variety of science cases. At the moment, line fitting in \gleam does not take into account line ratio predictions from radiative modeling and, as such, allows for any ratio between spectral lines. The option for a tighter coupling between line ratios could be desirable for specific science cases or, for example, low S/N regimes.

Another direction for development would be to investigate other back-ends for performing the line fitting, e.g.\ \texttt{astropy.modeling} from Astropy, which was not available at the time the main code development was occurring for \gleam. This approach would enable a closer integration with the Astropy suite of packages.

Further, as mentioned in Section~\ref{sec:examples}, we tested \gleam on a variety of optical and infrared observations. In the near future, we will test its robustness when applied to data at other wavelengths, such as radio (e.g.\ focusing on H\textsc{i} observations) and sub-mm observations (e.g.\ molecular and atomic lines, especially at high redshift).

\acknowledgments
The authors are grateful to the referee for their constructive suggestions, which improved the paper. Andra Stroe gratefully acknowledges the support of a Clay Fellowship. \gleam heavily relies on a number of scientific Python dependencies, including Astropy, LMFIT, Matplotlib, and NumPy. Its development makes use of other packages, tools, and services, including git, Poetry, Mypy, Black, Colorama, pydantic, Click, PyYAML, GitHub, and VSCode. In testing the software, we made use of observations obtained with the International Gemini Observatory, with ESO Telescopes at the La Silla Paranal Observatory, with the William Herschel Telescope, with the W.M. Keck Observatory, and with the MMT Observatory.
\vspace{5mm}
\software{
  \gleam \citep{stroe_gleam},
  Matplotlib \citep{matplotlib},
  Astropy \citep{2013A&A...558A..33A},
  LMFIT \citep{lmfit},
  Numpy \citep{harris2020array}
}

\bibliography{gleam}{}
\bibliographystyle{aasjournal}

\end{document}